\documentclass[12pt]{article}
\usepackage{epsfig}
\usepackage{amsfonts}

\def\laq{~\raise 0.4ex\hbox{$<$}\kern -0.8em\lower 0.62
ex\hbox{$\sim$}~}
\def\gaq{~\raise 0.4ex\hbox{$>$}\kern -0.7em\lower 0.62
ex\hbox{$\sim$}~}

\def\beq{\begin{equation}}
\def\eeq{\end{equation}}
\def\bea{\begin{eqnarray}}
\def\eea{\end{eqnarray}}

\def \la {\lambda}
\def \La {\Lambda}
\def \Da {\Delta}

\def \ga {\gamma}

\def \r {\rho}
\def \om {\omega}
\def \Om {\Omega}
\def \noi {\noindent}

\begin{document}
\begin{titlepage}

\begin{flushright}
BA-TH/694-15\\
\end{flushright}

\vspace{0.3 cm}

\begin{center}

\huge{Cosmology and short-distance gravity}

\vspace{1cm}

\large{M. Gasperini}

\bigskip
\normalsize

{\sl Dipartimento di Fisica,
Universit\`a di Bari, \\
Via G. Amendola 173, 70126 Bari, Italy\\
and\\
Istituto Nazionale di Fisica Nucleare, Sezione di Bari, Bari, Italy \\
\vspace{0.3cm}
E-mail: {\tt gasperini@ba.infn.it}

\vspace{0.3cm}
Date: {\rm 14 February 2015}}


\begin{abstract}
\noi
If the observed dark-energy density $\r_\La$ is interpreted as the net contribution of the energy density of the vacuum, $\r_\La\equiv \r_V \sim M_V^4$, and the corresponding vacuum length scale $\la_V = M_V^{-1}$ as the cutoff scale controlling the low-energy, effective field-theory limit of gravity, it follows that  the conventional cosmological  scenario based on the effective gravitational equations may be valid only up to the Tev energy scale. Such a possibility would be strongly disfavored by the existence of a relic background of primordial gravitational radiation of intensity compatible with present (or near future) experimental sensitivities.
\end{abstract}
\end{center}

\bigskip
\begin{center}
---------------------------------------------\\
\vspace {5 mm}
Essay written for the {\em 2015 
Awards for Essays on Gravitation,}\\
{  (Gravity Research Foundation, Wellesley Hills, MA, 02481-0004)}\\
 and awarded with {\em ``Honorable Mention"}\\
\vspace{0.7cm}
To appear in the  {\bf October 2015 Special Issue of Int. J. Mod. Phys. D}
\end{center}

\end{titlepage}

\newpage
\parskip 0.2cm

As is well known, recent astrophysical observations suggest the existence of a cosmological constant -- or a dark-energy component of the cosmic fluid-- with an energy density $\r_\La$ given by \cite{1}:
\beq
\r_\La \simeq 0.7 \,\r_c = 0.7 \times 3 \,H_0^2 M_P^2 \simeq 2.5 \times 10^{-47} \,{\rm GeV}^4.
\label{1}
\eeq
Here $\r_c$ s the critical density, $H_0$ the present value of the Hubble parameter, and $M_P= (8\pi G)^{-1/2}$ the reduced Planck mass. If such a cosmic energy density is interpreted as the gravitational contribution of the quantum (zero-point) fluctuations of the vacuum (see e.g. \cite{2}),
\beq
\r_\La =\r_V \equiv{1\over (2 \pi)^3} \int^{M_V} d^3k \, {1\over 2} \sqrt{k^2+m^2} \simeq {1 \over 16 \pi^2} \,M_V^4,
\label{2}
\eeq
and if the corresponding vacuum energy scale, $M_V$, is compared with the value of the more ``natural" Planck scale, $M_P \simeq 2.4 \times 10^{18}$ GeV, we find a discrepancy of about $30$ orders of magnitude:
\beq
M_V \simeq 8 \times 10^{-3} \,{\rm eV},  ~~~~~~~~~~~~ 
M_P \simeq 0.3 \times 10^{30}\, M_V.
\label{3}
\eeq

Should we interpret this fact as a signal that the effective ultraviolet cutoff of the gravitational interaction {\em is not} $M_P$ (as naively guessed) but, instead, it is given by the scale $M_V$ phenomenologically discovered? In other words, should we take seriously the possibility that the effective field-theory description of gravity we are currently using may undergo drastic (even quantum) modifications {\em not below the Planck length} $\la_P =M_P^{-1}$ but, instead, below the length scale $\la_V=M_V^{-1} \simeq 2 \times 10^{-3}$ cm? It should be noted, in this connection, that the direct  tests of the gravitational force are currently confirming the standard  theory only {above} distances of about $2 \times 10^{-2}$ cm (see e.g. \cite{3} for a review).

The possibility of an ``anomalous" gravitational scale has been suggested, for instance, by the so-called  model of ``fat gravitons" \cite{4}, where the predicted deviations from the standard gravitational laws at short (sub-millimeter) scales of distance are associated not to the presence of new particles or extra dimensions, but just to the existence of an effective cutoff much larger than  $\la_P$. It has been recently suggested, also, that a fundamental (quantum) model of gravity should be characterized, in general, by two mass scales \cite{5}:  a coupling scale $M_*$ (controlling the strength of the gravitational coupling) and a cutoff scale $M_{**}$ (controlling the effective theory limit),  different from each other and both different, in principle, from the Planck mass $M_P$.

We are thus motivated to ask the following question: what changes might we expect in the usual cosmological picture if the conventional limit of the effective, low-energy gravitational regime (the length scale $\la_P$) were to be replaced by the new phenomenological limiting scale $\la_V$? 

Let us consider a model of Universe based on the standard cosmological equations, supposing (for simplicity) that the strength of the gravitational coupling is still controlled by the Newton constant $G=(8 \pi M_P^2)^{-1}$: the total energy density $\r$ and the Hubble parameter $H$ are then related by the condition $\r= 3 H^2 M_P^2$. One usually assumes that such an effective model can be extrapolated up to the limiting curvature scale (or Hubble scale) $H \simeq M_P$, corresponding to the Planckian  density $3 M_P^4 \equiv \r_P$. If, instead, the limiting scale  is set by the condition $H_{\rm max} \simeq M_V$, it follows that we can apply the standard cosmological equations only up to a maximal energy  density, $\r_{\rm max}$, of the order of $3 M_V^2 M_P^2$. 

This limiting density is much smaller than the Planck density ($\r_{\rm max}\simeq 10^{-59}\r_P$),  yet extremely larger than the present  energy density given in  Eqs. (\ref{1}) and (\ref{2}), namely $\r_{\rm max}\simeq 3 \times 16 \pi^2 \times 10^{59} \r_V$. More precisely, we find
\beq
\r_{\rm max}= 3\, M_V^2 M_P^2 \simeq 1.1 \times 10^{15}\, {\rm Gev}^4,
\label{4}
\eeq
corresponding to an energy scale $E_{\rm max} =\r_{\rm max}^{1/4} \simeq 6$ TeV. We can easily obtain also the temperature associated to this scale by using again the standard cosmological equations (which can be applied for $\r \laq \r_{\rm max}$), and considering a radiation-dominated model of Universe in thermal equilibrium  at a temperature $T_{\rm max}$, such that:
\beq
\r_{\rm max}= {\pi^2\over 30}N_{\rm eff} T_{\rm max}^4.
\label{5}
\eeq
Using for the effective number of degrees of freedom the value  $N_{\rm eff} \sim 10^2$, typical of the Standard Model in the effective field-theory limit, we  then obtain $T_{\rm max} \simeq 2.4$ TeV. 

Given the above values of $E_{\rm max}$ and $T_{\rm max}$ it follows that there are no problems for the standard nucleosynthesis scenario, expected to occur at the Mev  scale. A maximal energy  of a few TeV is also compatible with appropriate electroweak mechanisms for Leptogenesis and Baryogenesis (see e.g. \cite{6,7}).
But what might happen to the Universe at the limiting epoch characterized by $H=H_{\rm max}$ and $\r=\r_{\rm max}$?
 
It is amusing to note, first of all, that the energy scale associated to this epoch ($E_{\rm max}\sim 6$ TeV) is still a possible candidate for the energy scale above which supersymmetry is  possibly restored -- at least according to ``constrained'' (or non-minimal)  versions of the Supersymmetric Standard Model (see e.g. \cite{8}). Aside from this, which in our context appears to be only a numerical coincidence\footnote{However, the coincidence of these two scales might acquire a dynamical explanation in the higher-dimensional context of supersymmetric brane-world cosmology \cite{9}.}, it is important to stress that inflation should necessarily occur at a curvature (or Hubble) scale $H_{\rm inf}$ not higher than the limiting curvature  scale $H_{\rm max} \simeq M_V$, in order to be described by  the standard equations of the effective gravitational field theory. This means, in other words, inflation at an energy scale {\em not higher than the {\rm TeV} scale}. 

It should be recalled, in this connection, that there are possible problems with models of inflation at the TeV scale based on the conventional (and even hybrid) slow-roll mechanism \cite{10} (see however \cite{6}). On the other hand, a phase of TeV-scale inflation like the one suggested by the above arguments has a number of interesting features. 

For instance, the value of the ``e-folding'' parameter $\cal N$ required for a successful inflationary scenario is much smaller than in the case of high-energy inflation. By applying the standard arguments, and calling $T_0$ and $T_{\rm eq}$ the temperature of the cosmic radiation at the present epoch and at the equality epoch, respectively, one finds indeed that the inflationary phase must satisfy the condition:
\beq
{\cal N} \gaq \ln \left(T_{\rm max}\over T_{\rm eq}\right)+{1\over 2} \left(T_{\rm eq}\over T_0\right).
\label{6}
\eeq
Using for $T_0$, $T_{\rm eq}$ the values given by the current experimental results \cite{1}, and imposing $T_{\rm max} \sim 1$ TeV, we then obtain
\beq
{\cal N} \gaq 32.
\label{7}
\eeq
It follows, in particular, that  the so-called ``trans-Planckian problem"  affecting the conventional inflationary scenario (see e.g. \cite{11}) could completely disappear in this context, provided the duration of the phase of Tev scale inflation is bounded by 
\beq
{\cal N} \laq {\cal N}_{\rm max} = \ln \left( \la_V \over \la_P \right) \simeq 68.
\label{8}
\eeq

In fact, only for ${\cal N} >{\cal N}_{\rm max}$ the quantum fluctuations with trans-Planckian wavelength $\la < \la_P$ have enough time to ```exit" from the horizon during inflation -- namely, are able to cross the inflationary horizon $H^{-1}_{\rm inf} = H^{-1}_{\rm max} \simeq M_V^{-1} = \la_V$ -- and can possibly contribute to the cosmic inhomogeneities that we are currently observing. 

Another important property of an inflationary phase at the Hubble scale $H_{\rm inf}= M_V$ is the associated reheating temperature, $T_{\rm rh}$, whose maximal value is given exactly by Eqs. (\ref{4}) and (\ref{5}), namely $T_{\rm rh} \laq T_{\rm max} \simeq 2.4$ TeV. As already stressed, such a temperature scale may be automatically compatible with Leptogenesis and Baryogenesis, in spite of the fact that the corresponding curvature scale is considerably lowered with respect to the one of the conventional inflationary models. 

It may be appropriate, at this point, to compare the picture we are sketching in this paper with the cosmological scenario recently discussed in \cite{5}, characterized by two gravitational scales: the short-distance cutoff $M_{**}$ and the effective coupling strength $M_*$ .

In our context, a possible model of inflation at the TeV energy scale (namely, a model with  $\r^{1/4}_{\rm inf} \simeq \r^{1/4}_{\rm max} \sim$ TeV) is obtained by lowering the cutoff scale (i.e. imposing $M_{**}= M_V \ll M_P$), while the  gravitational coupling is kept fixed  at $M_*=M_P$. In the context of \cite{5}, on the contrary, TeV-scale inflation is possibly implemented by lowering the coupling scale (i.e. by choosing $M_{*} \ll M_P$), at fixed cutoff $M_{**}\simeq M_P$. As a consequence, there are  important phenomenological differences between these two scenarios. 

Within the approach of \cite{5}, in particular, it is possible to formulate consistent models of TeV-scale inflation where the tensor fluctuations (i.e. the gravitational waves) of primordial origin are still a significant fraction of the observed spectrum of cosmological perturbations. This is obtained, however, at the price of introducing a huge number $N$ of different particle species, universally coupled to gravity with coupling strength controlled by $M_*$, and with masses lower than this scale.
For instance, denoting with $r_*$ the usual tensor-to-scalar ratio of the primordial power spectra, and considering a model of inflation at the energy scale $\r^{1/4}_{\rm inf} \sim 10$ TeV, one can obtain a significant fraction of tensor perturbations, say $r_* \gaq 0.01$, provided the model contains a number $N \gaq 10^{24}$ of universally coupled degrees of freedom, with masses not higher than the $10$ TeV scale (presumably, towers of states of higher-dimensional and/or string theory origin, as discussed in \cite{5}). 

According to the picture suggested in this paper, on the contrary, a background of tensor perturbations produced at the TeV scale, and its possible contribution to the observed properties of the CMB radiation (such as anisotropy, polarization, and so on), turns out to be characterized by an intensity (and a value of the parameter $r_*$) which is always completely negligible.

Let us consider, in fact, a primordial background of relic gravitational radiation amplified by inflation, whose spectral distribution ranges today from the minimal frequency $\om_0$ associated to the present Hubble scale, $\om_0 \simeq H_0 \sim 2 \times10^{-18}$ Hz, to the maximal (cutoff) frequency $\om_1$, determined by the inflation scale as $\om_1 \sim (H_{\rm inf}/M_P)^{1/2} \times 10^{11}$ Hz (see e.g. \cite{12}). In our case $H_{\rm inf} \simeq M_V$, so that $\om_1 \sim 10^{-4}$ Hz.
If the tensor spectrum is flat, or decreasing in frequency (as predicted  for instance by models of slow-roll inflation), then the peak intensity is reached at the lowest-frequency end of the spectrum, and the corresponding  fraction of critical energy density, in our case, is presently given by
\beq
\Om_{\rm gw} (\om_0, t_0) \laq \left(H_{\rm inf}\over M_P \right)^2 \simeq
 \left(M_V\over M_P \right)^2  \sim 10^{-59}.
 \label{9}
 \eeq
 As a comparison, the tensor amplitude needed to produce currently observable effects on the CMB radiation, at large scales, is typically of the order of $\Om_{\rm gw} (\om_0, t_0) \sim 10^{-11}$. 
 
 If the phase of TeV-scale inflation leads instead  to a less conventional ``blue" (i.e. growing) tensor spectrum, then the tensor amplitude is even lower: the peak intensity is reached at the high-frequency end of the spectrum, the cutoff $\om_1$ (see e.g. \cite{12}), and the corresponding fraction of critical  density is given, in this case, by
\beq
\Om_{\rm gw} (\om_1, t_0) \sim \Om_\ga (t_0) \left(H_{\rm inf}\over M_P \right)^2 \simeq \Om_\ga (t_0) 
 \left(M_V\over M_P \right)^2  \sim 5 \times 10^{-64},
 \label{10}
 \eeq
where $\Om_\ga (t_0) \sim 5 \times 10^{-5}$ is the present value of the CMB energy density, in critical units \cite{1}. In any case, the background of tensor perturbations associated to the model of Tev-scale inflation considered in this paper is too low to produce any detectable effect.

The same conclusion can be expressed in terms of the parameter $r_*$, controlling the relative intensity of the tensor spectrum. For instance, if we apply the conventional slow-roll mechanism at the inflation scale $H_{\rm inf} \simeq M_V$, we obtain a power spectrum of scalar perturbations $\Da_{\cal R}^2$ which can be parametrized as follows:
\beq
\Da_{\cal R}^2 \simeq {2\over \pi^2} \left( H_{\rm inf}^2\over r_* M_P^2 \right) \simeq
{2\over \pi^2} \left(M_V \over M_P \right)^2 {1\over r_*} \equiv A_s \left( \om \over \om_*\right)^{n_s-1}.
\label{11}
\eeq
Here $A_s$ is the spectral amplitude at a given frequency scale $\om_*$, and $n_s$ the scalar spectral index. Using the recent experimental results \cite{1} specified at the reference scale $\om_*= 0.05 \,{\rm Mpc}^{-1}$, namely $A_s \simeq 2.2 \times 10^{-9}$, we immediately obtain, at that scale:
\beq
r_* \simeq {2\over \pi^2 A_s} \left(M_V \over M_P \right)^2 \simeq 10^{-51}.
\label{12}
\eeq
This value, well below the current experimental sensitivities \cite{1}, confirms the basic differences between the scenario considered in this paper and other possible models of TeV-scale inflation (like, for instance, those  based on the theoretical scheme discussed in \cite{5}).

We may thus conclude that any direct/indirect detection of primordial tensor perturbations would severely disprove the possibility that the observed dark-energy density be associated to a new fundamental scale $M_V$, replacing the Planck scale in controlling the effective field-theory limit of gravity. Conversely, the absence of any detectable signal from a tensor component of the primordial perturbation spectrum may be regarded as an indirect support to this possibility.

\section*{Acknowledgements}

This work is supported in part by MIUR, under grant no. 2012CPPYP7 
(PRIN 2012), and by INFN,  under the program TAsP ({Theoretical Astroparticle Physics}).

\vspace{1 cm}


\end{document}